\documentclass{epsconf}
\usepackage{graphicx}
\usepackage{wrapfig}
\usepackage{amsmath,amssymb}
\newcommand{\ie}{\emph{i.e., }}

\newcommand{\figref}[1]{Fig.\ref{#1}}
\newcommand{\p}{\partial}
\newcommand{\etab}{\bar{\eta}}
\newcommand{\rb}{\bar{r}}

\newcommand{\omb}{\bar{\omega}}
\newcommand{\omp}{\omega_p}

\title{Hamiltonian bump-on-tail model:\\interpretation of EP/AE interaction}
\author{\underline{N. Carlevaro}$^{1,2}$, G. Montani$^{1,3}$, X. Wang$^{4}$, F. Zonca$^{1}$}
\institute{$^1$ENEA - FSN-FUSPHY-TSM, R.C. Frascati (Italy). $^2$L.T. Calcoli, Merate (Italy).\\
$^3$Physics Department, ``Sapienza'' University of Rome (Italy).\\
$^4$ Max Planck Institute for Theoretical Physics, Garching (German).}

\begin{document}
\maketitle

\paragraph{Abstract}
The Bump-on-Tail (BoT) model is often adopted to characterize the non-linear interaction between fast ions and Alfv\'en Eigenmodes (AEs). A multi-beam Hamiltonian approach to the BoT model is tested here as paradigm for the description of these phenomena. 

\paragraph{Introduction}
In this work, we reproduce the non-linear dynamics of a single beta-induced Alfv\'n Eigenmode (BAE) resonance treated in \cite{WB16}, with a one-dimensional (1D) $N$-body description of the beam-plasma system (BPS) instability \cite{JPP,Entropy} in the presence of an isolated resonant mode. For a single toroidal number and constant frequency, the quantity $C=\omega_{BAE} P_\phi-n_{BAE}E$ (where $P_\phi$ and $E$ are the particle toroidal angular momentum and energy, respectively, while $n_{BAE}$ denotes the toroidal mode number and $\omega_{BAE}$ the mode frequency), and the magnetic moment $\mu$ are constants of the particle (perturbed) motion. Cutting the energetic particle (EP) phase space into slices of given $\mu$ and $C$, particles remain, thus, in the same slice during the whole evolution: the wave-particle power exchanges within different slices are then independent of each other. The mode evolution, however, is consistent with the presence of all the EP phase space slices (for details on Hamiltonian mapping technique, see \cite{B14}).

A proper dimensional reduction of the phase-space dynamics is at the ground of the possibility to use the BoT paradigm in this framework. In other words, by selecting constants of motion for the particle dynamics, we are able to reduce the distribution function evolution to a 1D non-autonomous problem. For an assigned initial subdivision of the EP phase space according to a set of integrals of motion (here $C$ and $\mu$), we can map each independent slice into and equivalent 1D BoT problem. Such a prescription is a necessary ingredient provided, in general, by a multi-dimensional (linear) numerical analysis, to be complemented by the mapping to the equivalent BoT problem described below.

\paragraph{Theoretical Framework} The mapping between the reduced radial profile ($r$) and the BPS velocity ($v$) space is a one-to-one link between the two corresponding independent variables. It is derived from the resonance condition\footnote{Following \cite{WB16}, the EP/BAE system is characterized by toroidal mode number $n_{BAE}=2$ and the poloidal harmonic $m_{BAE}=4$. The normalized Tokamak radius reads $\rb=r/a$ ($a$ denotes the minor radius), while frequencies are normalized as $\omb=\omega/\omega_{A0}$ (with $\omega_{A0}=v_{A0}/R_0$, where $v_{A0}$ is the Alfv\'en speed at the magnetic axis and $R_0$ the major radius). The aspect ratio is set as $R_0/a=10$ and fast ions (hot) velocity is assumed as $v_H=0.3v_{A0}$. At the same time, the BPS consists in a background plasma with constant particle density $n_p$ and beams with total number density $n_B$. The plasma is assumed cold, thus the dielectric function reads $\epsilon=1-\omp^{2}/\omega^{2}$ (the plasma frequency is $\omp^{2}=4\pi n_p e^2/m_e$). The periodicity length of the system is indicated as $L$, thus the resonant wave-number can be normalized as $\ell_{res}=k_{res}(2\pi/L)^{-1}$.}
$\omb_{res}(\rb)-\omb_{res}(\rb_{res})=k_{res}(v-v_{res})/\omega_{A0}$ (where $v_{res}$ is the resonant velocity of the BPS), by defining a local map trough the expansion of $\omb_{res}$ near $\rb_{res}$ (the resonant normalized radius) as $\omb_{res}-\omb_{res}(\rb_{res})=(\rb-\rb_{res})\p_{\rb}\omb_{res}|_{\rb_{res}}\equiv(\rb-\rb_{res})\omb_{res}'$:
\begin{align}
\rb=\rb_{res}+k_{res}(v-v_{res})/(\omb_{res}'\omega_{A0})\;.
\end{align}

\begin{figure}[ht!]
\centering
\includegraphics[width=.35\textwidth,clip]{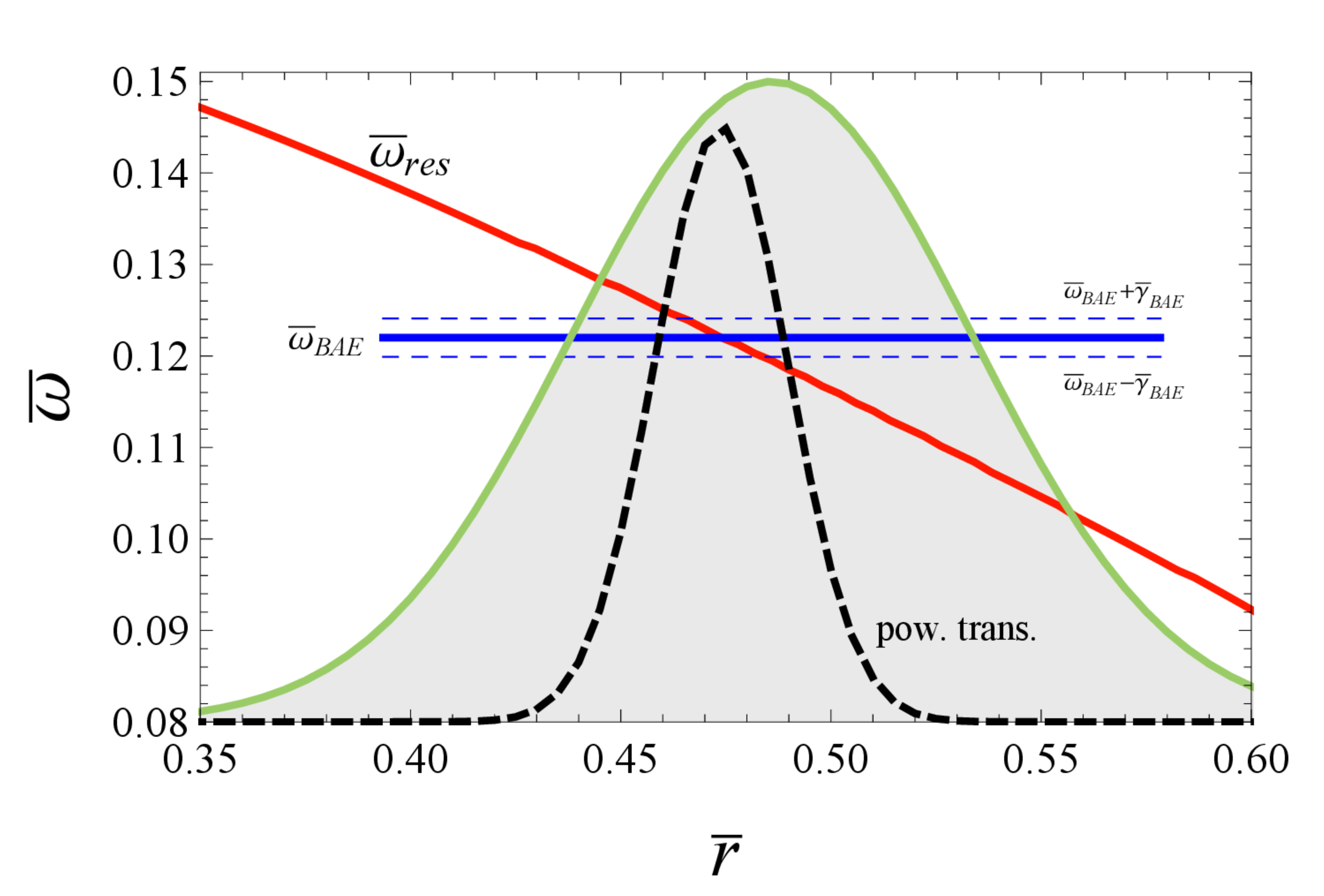}\quad
\includegraphics[width=.345\textwidth,clip]{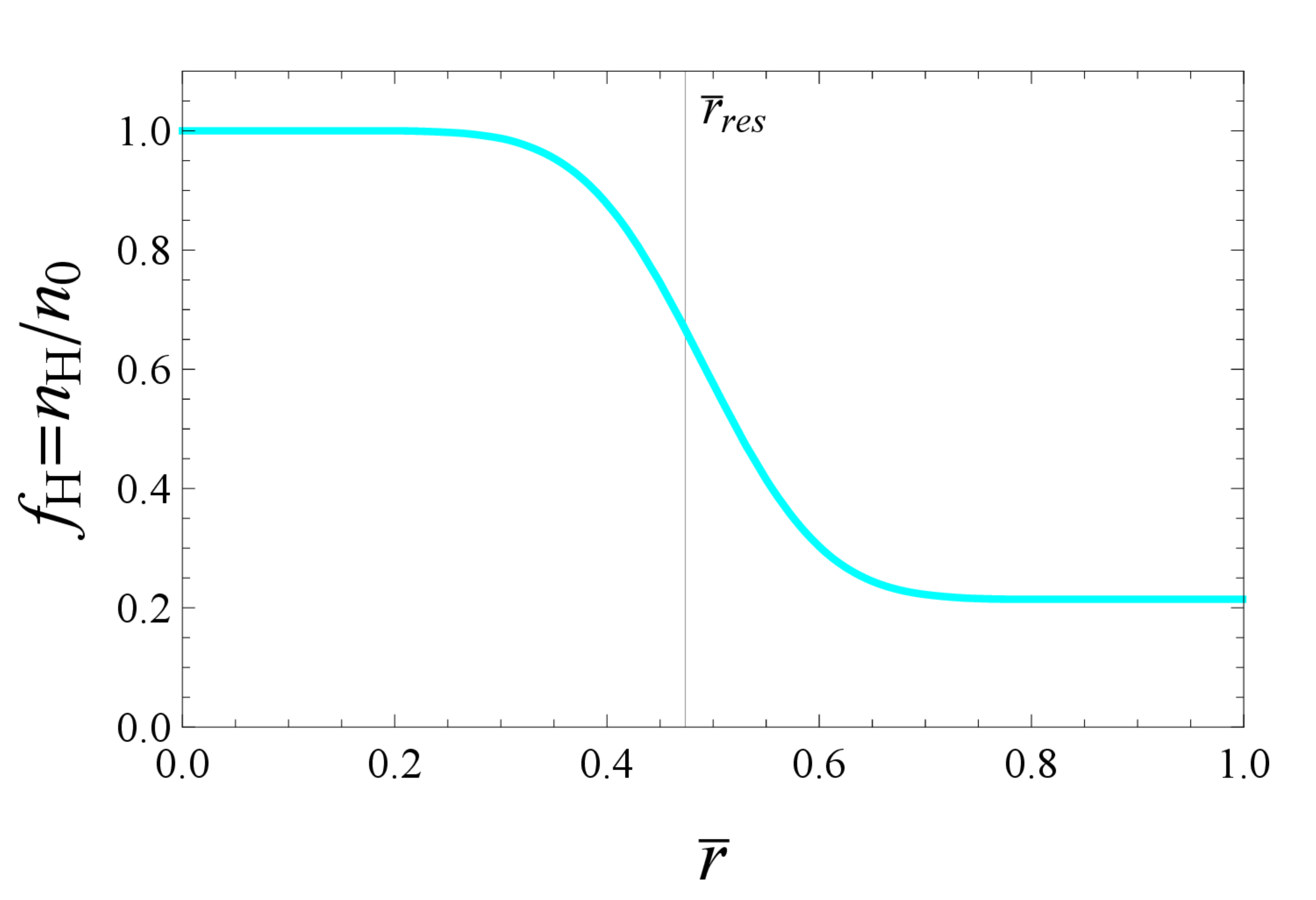}\vspace{-5mm}
\caption{\it \small 
Left-hand panel: Resonance structure and frequencies (indicated in the plot), mode structure (filled green) and effective power transfer (dashed black).\;\;Right-hand panel: EP density radial profile.\label{fig1}}
\end{figure}

The instability drive $\gamma_L$ for the BPS is obtained from the normalized beam distribution function $\hat{f}_B=f_B/n_B$ as
\begin{align}
\gamma_L/\omega_0=\pi(\omega_0/k_{res})^{2}\etab^{3}\p_v\hat{f}_B\big|_{v_{res}}\;,
\end{align}
where $\etab\equiv(n_B/2n_p)^{1/3}$ and $\omega_0=\omp$ is the corresponding Langmuir wave frequency. Moreover, for the considered resonant mode, we assume the following resonance condition $k_{res}v_{res}=\omega_0$. Here, we impose the proper BPS drive in order to recover the BAE linear growth rate given in \cite{WB16} (specified for a fixed fast-ion density): $\bar{\gamma}_{BAE}/\omb_{BAE}=\gamma_L/\omega_0$ with $\omega_0=\omp=\omb_{BAE}\omega_{A0}$.
Imposing now the constraints on the normalized radius (fixing a reference frame for the velocity space), \ie $\rb_{min}=0\mapsto v_{Max}$,\; $\rb_{Max}=1\mapsto v_{min}=0$,
and reproducing with $\hat{f}_B(v)$ the normalized EP radial profile $f_H(\rb)$ (right-hand panel of \figref{fig1}), we finally get
\begin{align}
\etab^3=\frac{\bar{\gamma}_{BAE}}{\omb_{BAE}}\;
\Big[\pi(1-\rb_{res})^{2}\frac{-\p_{\rb} f_H|_{\rb_{res}}}{\int_0^1d\rb f_H}\Big]^{-1}\;.
\end{align}

Following the reference case of \cite{WB16}, we now consider the dimensional reduced analysis for a given ``resonant'' slice characterized by the largest power exchange. We, thus, get (as shown in the left-hand panel of \figref{fig1}) the resonance condition $\rb_{res}=0.474$, with $\bar{\gamma}_{BAE}=0.0021$ and $\omb_{BAE}=0.122$.
We then obtain: $\etab=0.151$ and $\gamma_L=\bar{\gamma}_L\etab\omp$ with $\bar{\gamma}_L=0.114$. Using dimensionless velocities $v=\omp\etab(2\pi/L)^{-1}\;\nu/\ell_{res}$, the mapping can be recast as
\begin{align}
\nu=\nu_{res}-(\rb-\rb_{res})/A\;,\qquad\quad
A=\etab\omb_{BAE}/\omb_{res}'=-0.0823\;.
\end{align}

\begin{figure}[ht!]
\centering
\includegraphics[width=.344\textwidth,clip]{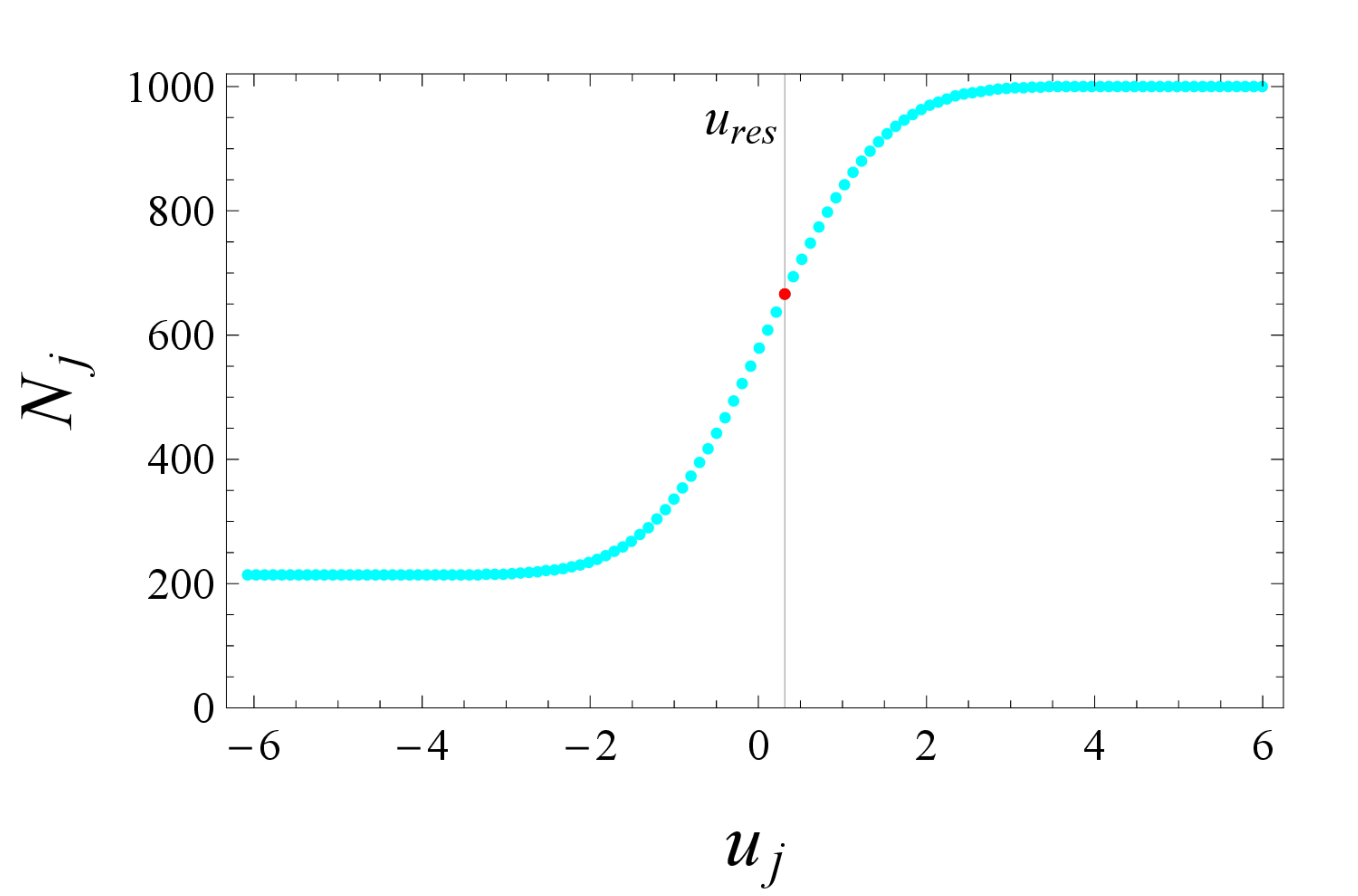}\quad
\includegraphics[width=.35\textwidth,clip]{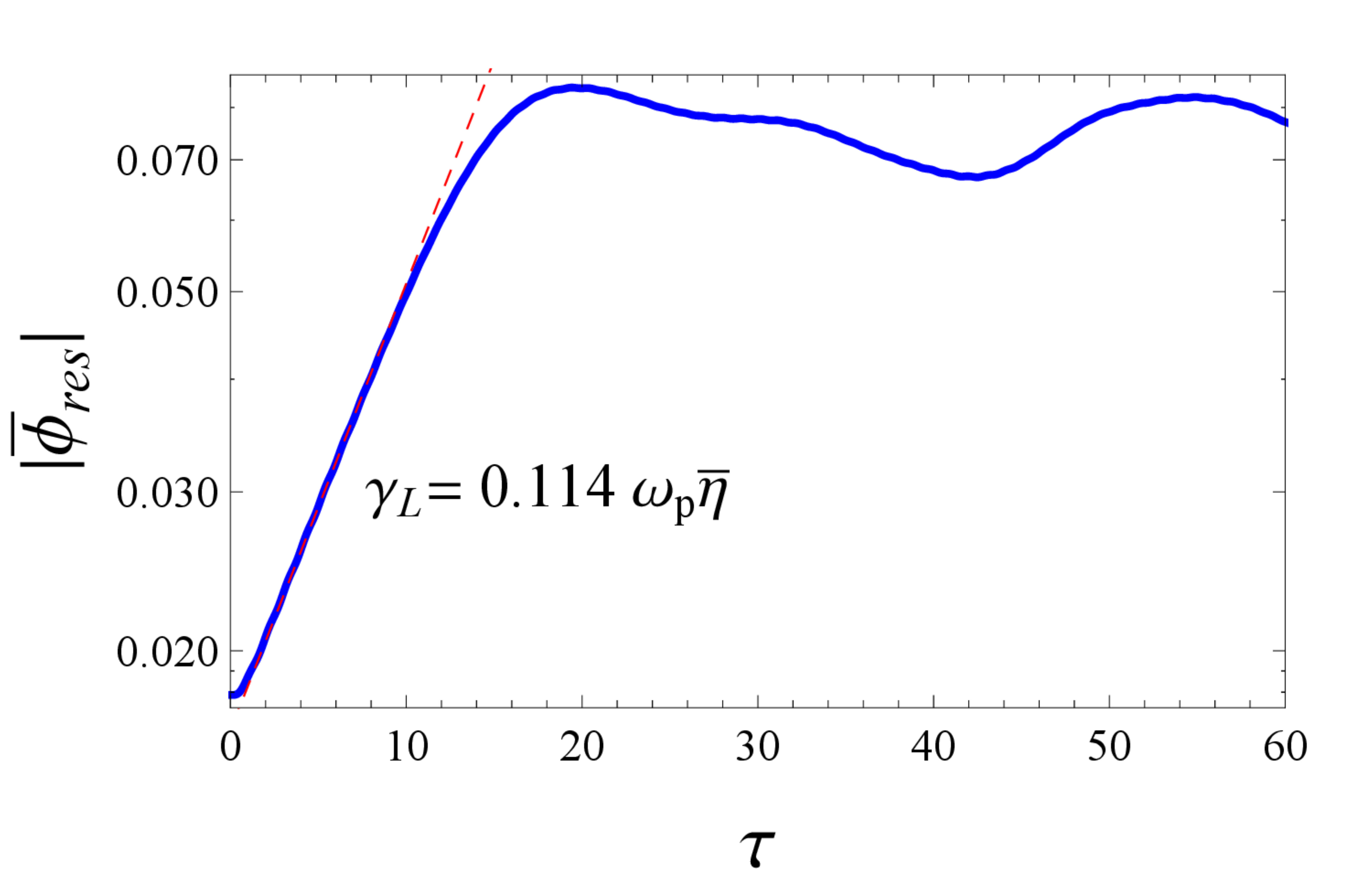}\vspace{-5mm}
\caption{\it \small Left-hand panel: Initial distribution of particles for the BPS.\;\; Right-hand panel: Langmuir mode evolution in Log scale and the line (dashed red) representing the initial exponential evolution.\label{fig2}}
\end{figure}

We now sample the fast-ion density radial profile $f_H(\rb)$ in $n=600$ ``beams'', and formally introduce the number of particles $N_j$ (with $j=1,\,...,\,n$), located at $\rb_j$, for the $N$-body simulation: we use $N=3.6\times10^5$ total particles. From the constraint $0\leqslant\rb\leqslant1$, using dimensionless velocities, we obtain $\nu_{res}=(\rb_{res}-1)/A$. For simplicity, we move to the reference frame of the average beam speed, $u=\nu/\ell_{res}-\langle\nu\rangle/\ell_{res}$, and arbitrarily fix the resonant normalized wave-number ($\ell_{res}=1$).
The velocity initial conditions of beam particles (left-hand panel of \figref{fig2}) are defined from the $\rb_j$-sampling using the mapping above, with the initial distribution defined by $N_j$. This system is evolved self-consistently in order to generate the dimensionless potential $\bar{\phi}_{res}$ (right-hand panel of \figref{fig2}): simulation results are consistent with the assumed $\gamma_L$ and correspond to an initial exponential evolution (in red in the figure) followed by mode saturation ($|\bar{\phi}_{res}|^{SAT}\simeq0.084$) and the consequent non-linear oscillation.

\begin{figure}[ht!]
\centering
\includegraphics[width=.35\textwidth,clip]{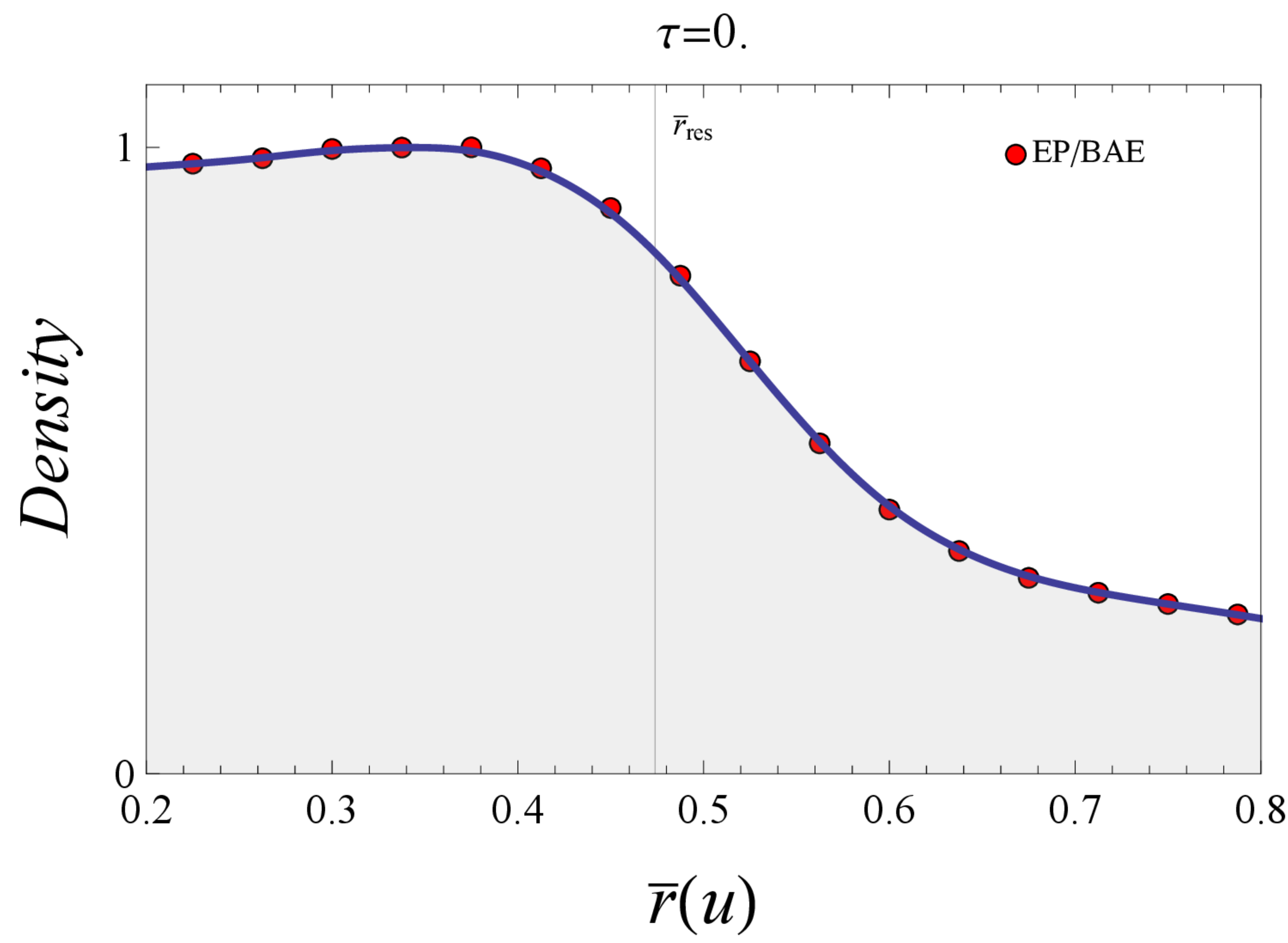}\quad
\includegraphics[width=.35\textwidth,clip]{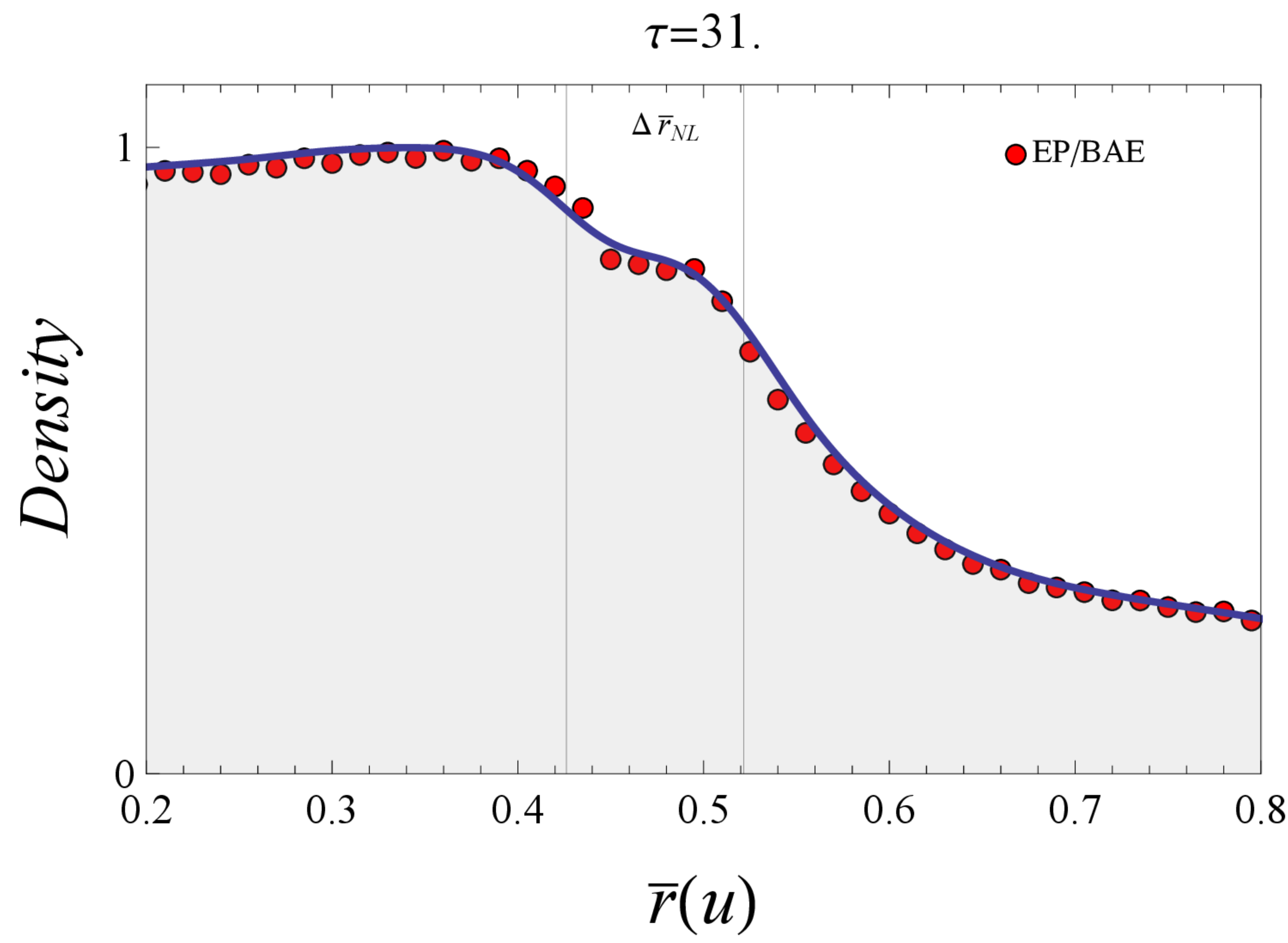}\vspace{-5mm}
\caption{\it \small Left-hand panel: Initial density distribution of test particles.\;\; Right-hand panel: Density profile around saturation. (Blue Line: BPS evolution mapped back to $\rb$ space. Red Bullet: data from \cite{WB16})\label{fig3}}
\end{figure}

\vspace{-5mm}
\paragraph{Numerical Analysis} Let us now address predictivity of the obtained numerical results on the reduced 1D radial profile evolution. A direct comparison between the self-consistent EP/BAE distribution function and that obtained from our BPS simulations is shown in  \figref{fig3}. The very good agreement of the two distribution functions is evident, demonstrating the reliability of the proposed mapping procedure. It is worth noting that the observed density flattening width is also in agreement with the BPS estimate of the non-linear velocity spread $\Delta u_{NL}\simeq\sqrt{4\bar{\phi}_{res}^{SAT}}$ (in the right-hand panel of \figref{fig3}, we indicate the mapped back value $\Delta\rb_{NL}$), suggesting a simple predictive model of this behavior.
\begin{figure}[ht!]
\centering
\includegraphics[width=.3\textwidth,clip]{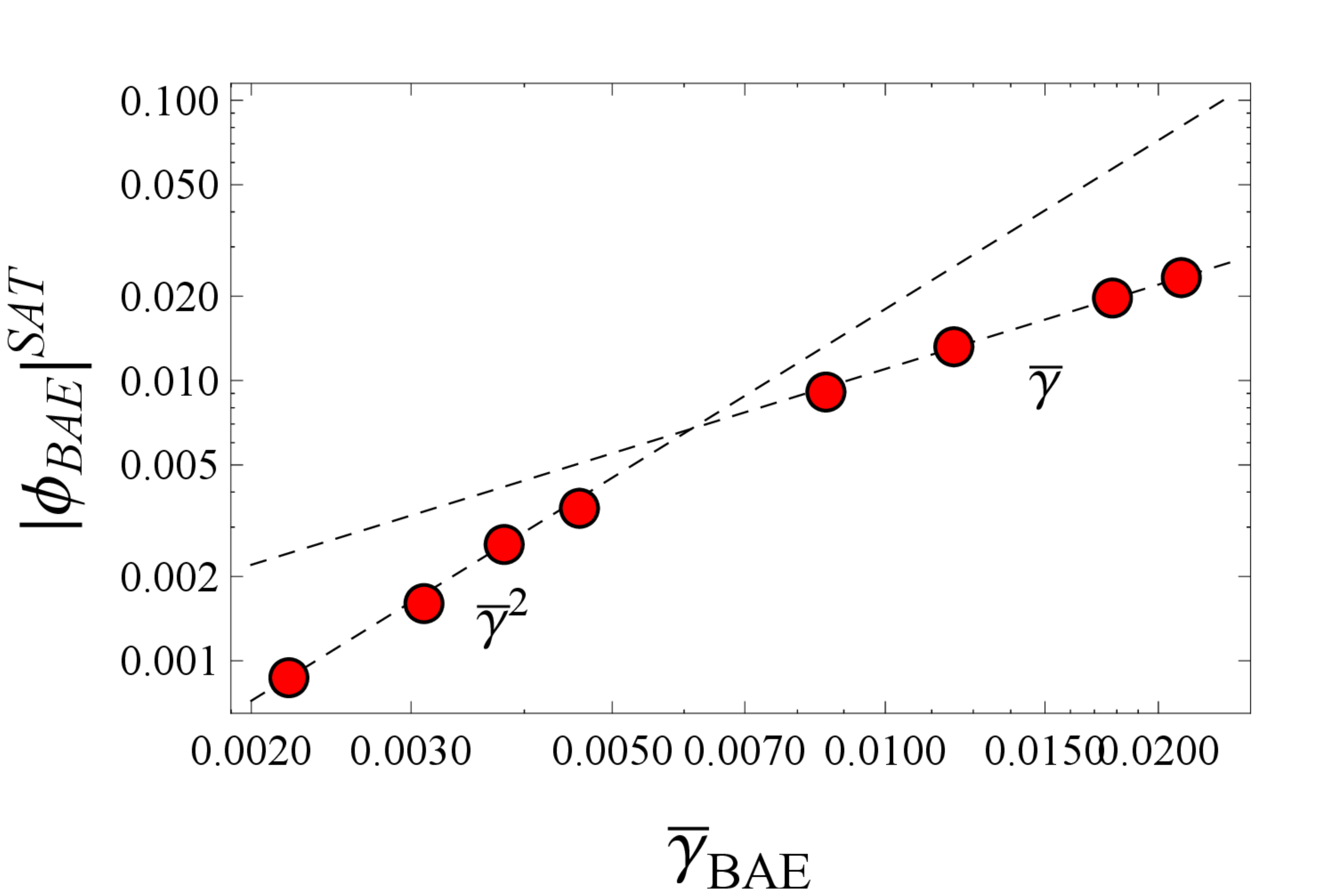}\quad
\includegraphics[width=.295\textwidth,clip]{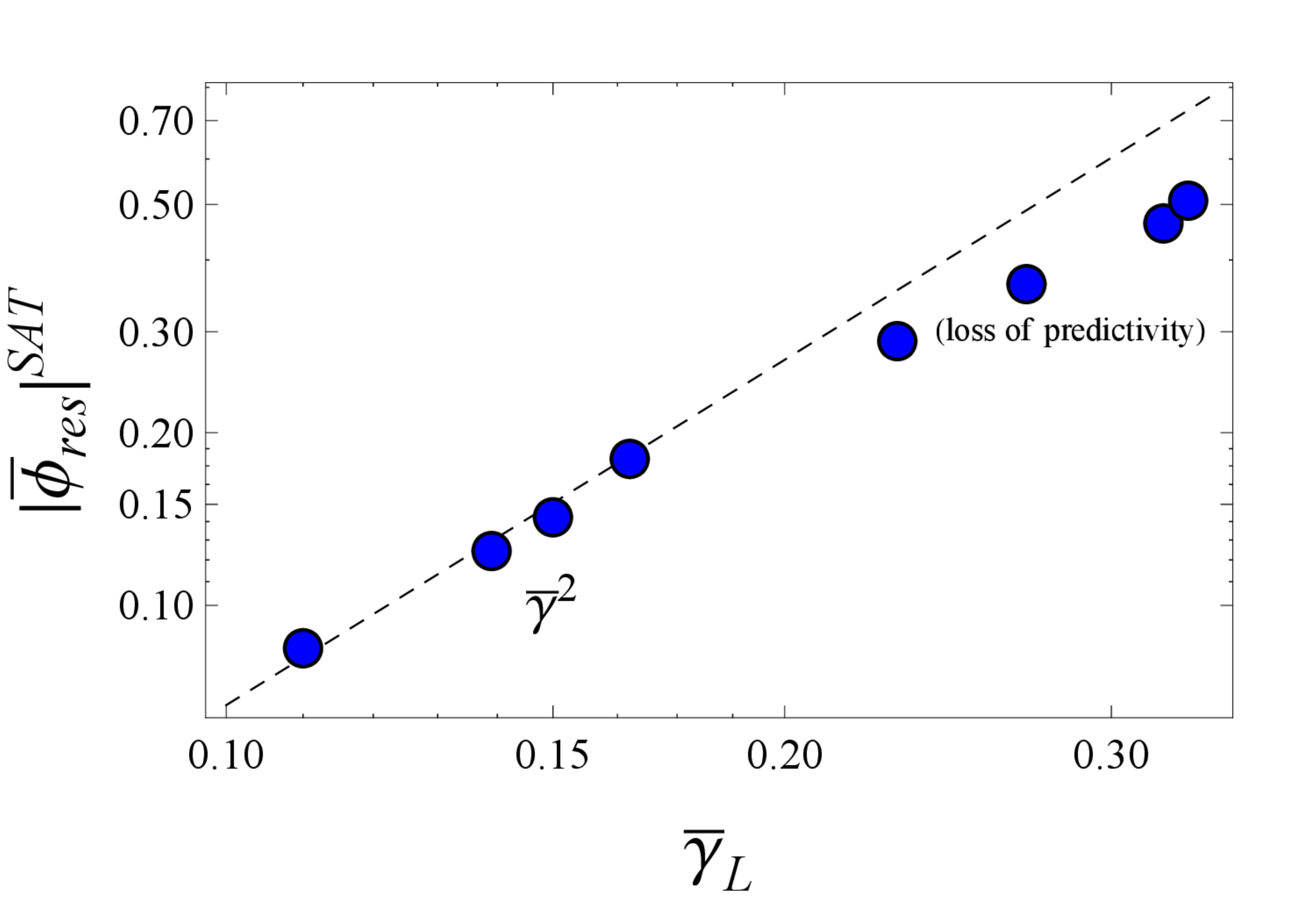}\quad
\includegraphics[width=.3\textwidth,clip]{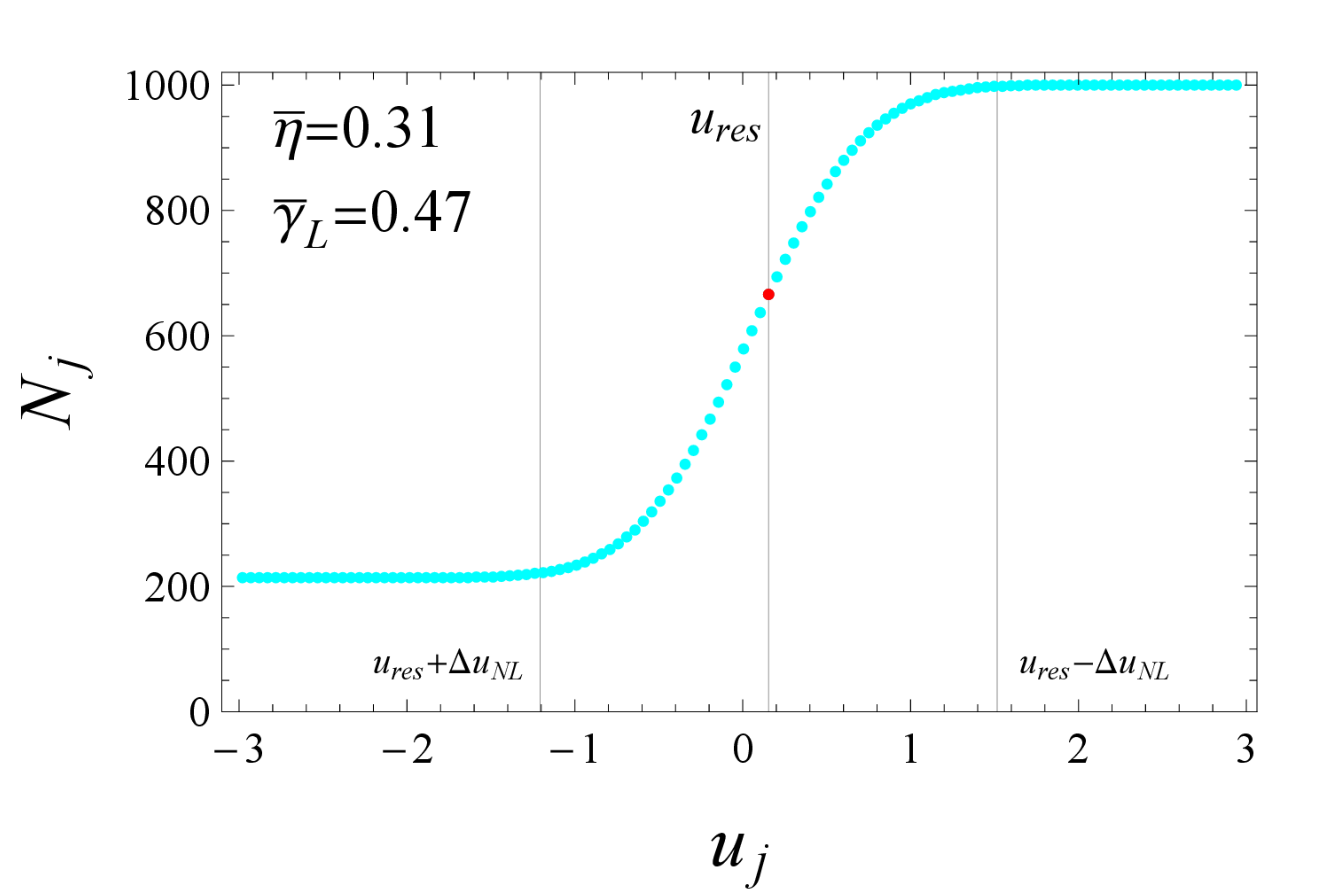}\vspace{-5mm}
\caption{\it \small Left-hand panel: Saturation amplitude of the scalar potential vs $\bar{\gamma}_{BAE}$. \;\; Center panel: Langmuir mode saturation level in the BPS for correspondingly different drive ($\bar{\gamma}_{BAE}/\omb_{BAE}=\gamma_L/\omp$).\;\; Right-hand panel: Initial distribution and non-linear velocity spread for the BPS, in the case of large drive. \label{fig4}}
\end{figure}
Finally, we observe how (see \figref{fig4}) the growth rate scaling with the mode saturation amplitude, for the EP/BAE system, is quadratic as far as the resonance width (power transfer region) is smaller than the mode structure. Otherwise, the behavior is linear. Analogously, the quadratic scaling is also recovered for the BPS system, while the deviation for large $\gamma_L$ values occurs when $\Delta u_{NL}$ becomes so large that flat regions of the initial distribution function are affected by nonlinear dynamics (as depicted in the right-hand panel of \figref{fig4}): in this limit the BPS model clearly fails.

\paragraph{Outlooks} The obtained results constitute the starting point for the investigation of more realistic cases of relevance for ITER with the present approach, \ie the analysis of multi resonance regimes for which different resonant regions overlap \cite{SL16}. Finally, two further conceptual questions must be properly addressed: \emph{(i)} properly accounting for the intrinsic multi-dimensional features in the reduction of the AE dynamics to the 1D BoT model; \emph{(ii)} introducing effective \emph{form factors} in order to model the finite mode structure and recover the linear $\gamma$ scaling of mode saturation by radial decoupling.

\end{document}